\documentclass[twocolumn,pra,superscriptaddress]{revtex4}
\usepackage[utf8]{inputenc}
\usepackage{amsmath}
\usepackage{amssymb}
\usepackage{bm}
\usepackage{graphicx}
\usepackage{color}

\begin{document}

\title{Configurational Mean-Field Reduced Transfer Matrix \\ Method for Ising Systems}

\author{Tuncer Kaya}\email{tkaya@yildiz.edu.tr}
\affiliation{Department of Physics, Yıldız Technical University,
	34220 Davutpaşa-Istanbul/Turkey}

\author{Başer Tambaş}\email{tambas19@itu.edu.tr}
\affiliation{Department of Physics, Istanbul Technical University,
	34469 Maslak-Istanbul/Turkey}

\begin{abstract}

A mean-field method for the hypercubic nearest-neighbor Ising system is introduced and applications to the method are demonstrated. The main idea of this work is to combine the Kadanoff's mean-field approach with the model presented by one of us previously. The mean-field approximation is introduced with the replacement of the central spin in Ising Hamiltonian with an average value of particular spin configuration, i.e, the approximation is taken into account within each configuration. This approximation is used in two different mean-field-type approaches. The first consideration is a pure-mean-field-type treatment in which all the neighboring spins are replaced with the assumed configurational average. The second consideration is introduced by the reduced transfer matrix method. The estimations of critical coupling values of the systems are evaluated both numerically and also analytically by the using of saddle point approximation. The analytical estimation of critical values in the first and second considerations are $ K_{c}=\frac{1}{z} $ and $ (z-2) K_{c}e^{2K_{c}} =1 $ respectively. Obviously, both of the considerations have some significant deviation from the exact treatment. In this work, we conclude that the method introduced here is more appropriate physical picture than self-consistent mean-field-type models, because the method introduced here does not presume the presence of the phase transition from the outset. Consequently, the introduced approach potentially makes our research very valuable mean-field-type picture for phase transition treatment.
\newline

\end{abstract}

\maketitle 

\section{Introduction}
One of the physically simplest models of magnetism is the nearest-neighbor Ising model \cite{Ising}. But, it has been solved exactly just a few simple cases. The exact solution in one dimension (1D) has been obtained by Ising. The important prediction of the 1D solution is that 1D Ising system exhibits no phase transition at a nonzero temperature. In two dimension (2D), the exact critical temperature for square lattice was estimated by Kramers and Wannier \cite{Kramers}. Shortly afterwards, Onsager \cite{Onsager} determined the free energy exactly for 2D. It has a long story, but we advise Baxter's recent paper \cite{Baxter} to get more information about the historical development of 2D Ising system. The main outcome of the solution is that there is spontaneous magnetization in 2D and the corresponding critical coupling strength $ K_{c} $ is equal to $ 0.44 $. This result is particularly important in that it is one of a few nontrivial example of phase transition that can be worked out with mathematical rigor. Other exact results were obtained for honeycomb and triangular lattices \cite{Domb,Baxterb}. For other lattices and higher dimensions, there is no exact solutions. However, various mean-field theories have been proposed to get some insight into the properties of the systems. We think it is important to mention that in the vicinity of a critical point, mean-field theories of Ising models in general breaks down due to the neglected statistical fluctuations. Therefore, it is necessary to resort to more sophisticated theories such as scaling and renormalization group \cite{Efrati} in order to study the detailed critical properties. On the other hand, mean-field-type theories are still important in that they are capable of predicting the general feature of the system over a wide range of the parameters with a relatively easy manner. In addition, estimations of critical coupling strengths of mean-field-type theories are fairly good when they are compared with the series or exact predictions. Therefore, a further improvement of the mean-field-type theory is still important and many attempts have been made in recent years \cite{Kaplan,Chamber,Zhur,Yama,Kayaa,Kayab}. \\

The mean-field-type treatment of Ising model has a long history. It is not possible to include all of the contribution in this paper. But, it is proper to mention a few of them. The mean-field approach developed by Weiss \cite{Weiss} was very successful at showing a nonzero transition critical temperature for Ising system. However, any self-consistent mean-field-type approximation assumes that there exist phase transition from the outset since any self-consistent approximation inevitable leads to a self-consistent critical relation. Of course, this point makes the physical relevance of any self-consistent approaches questionable. In the realm of this paper, we will try to consider this physical ambiguity. In order to clarify this point let us consider the Weiss mean-field approach. In Weiss's model, a given spin experiences both the external applied magnetic field and the effective field due to the other spins, which is expressed as $ h_{eff}=zJ\langle\sigma\rangle $, here $ z $, $ J $ and  $ \langle\sigma\rangle $  are the coordination number, coupling constant and average magnetization per spin respectively. The combination of the two fields then determine the response of the given spin or the average magnetization as $ \langle\sigma\rangle=\tanh(\beta Jz\langle\sigma\rangle +h) $. This self-consistent relation leads to well-known critical relation of Weiss,  $ K_{c}=\frac{1}{z} $ in the limit $ h\rightarrow0 $ and $ \langle\sigma\rangle\ll1 $. Notice that the prediction of the critical relation is a result of assumed effective field relation $ h_{eff} $ which depends on the average magnetization linearly and has to lead a critical point correlation inevitably. \\        

The other two most important mean-field-type approximations are the Bragg-Williams \cite{BW} and the Bethe-Peierls \cite{Bethe,Huang} models. Ignoring the possibility of local correlation between spins, the Bragg-Williams theory predicts that the critical coupling strength $ K_{c} $ is equal to $ \frac{1}{z} $, here $ z $ is the number of nearest-neighbor of any given site or coordination number. The main drawback of Weiss model and also Bragg-Williams approximation is due to the wrong prediction of spontaneous magnetization for 1D system, an impossibility not present in the exact solution. Taking into account the local correlation between spins approximately by the aid of heuristically introduced long range and short range order parameters, the Bethe-Peierls theory has a prediction of $ \frac{1}{2}\ln \left(\frac{z}{z-2}\right) $ for $ K_{c} $. Apparently, this result leads to an exact prediction for 1D Ising model. \\

Despite its known limitation, it is still important to consider how to improve the mean-field approach by including more correlations effects \cite{Honmura,HK,Wilsona,Wilsonb,Fisher}. Improvements in this respect \cite{Taggrat,Kroes} have been worked out by many methods as pointed out previously. The common focus of most of the models are to improve the estimations of the critical coupling strengths. However, the assumed approximations used in most of the mean-field-type approaches are somewhat unsatisfactory. For example, replacing the neighboring spin with its ensemble average turn out to be quiet uncontrolled approximation. In other words, neither the range of the validity nor the size of the omitted physics is not clear from this type of assumption. Therefore, we think that it is important to introduce a physically more relevant and clear mean-field model. For this purpose, we are going to mostly focus on the physical transparency and relevancy instead of the better estimations of critical coupling strengths. \\

In this paper, as an attempt to improve the physical relevancy of mean-field Ising model, we introduced a new approximate method, which we refer to as ``Configurational mean-field reduced transfer matrix'' theory based on Kadanoff's mean-field approach \cite{Kadanoff}. What we have mainly done in this paper, we combined Kadanoff's approach together with the ideas presented in the work written by one of us \cite{Kayab}. In this present approach, the central spin is taken into account within an average of each configuration. The neighboring spins are included exactly for each configuration. In the next section, all the central spins in all configurations are treated with the average. This consideration with the aid of saddle point approximation leads exactly the same estimation of Weiss, $ K_{c}=\frac{1}{z} $. Since this result predicts spontaneous magnetization for 1D system at finite temperature, the model extended with the use of previously introduced reduced transfer matrix model in the following section. The physical approach is more or less similar to the pure mean-field case, but the interaction between central spin and one of its neighbor is included exactly in the calculations. We have obtained the critical coupling relation as, $ (z-2) K_{c}e^{2K_{c}} =1 $, which is totally in agreement with the exact solution of 1D Ising model. In both of the following two section, some numerical calculations are also presented for the average magnetization per spin due to the some draw back of the saddle point approximation for $ K>K_{c} $. In the last section, some discussion is also presented.    

\section{Pure Mean-Field Model}
In general, mean-field-type theory is the simplest treatment of an interacting statistical mechanical system, apart from the approximation of ignoring the interaction all together. Often it is almost trivial to perform the mean-field theory calculation. In addition, the existence of phase transition is inserted inside the theory from the outset as in the case of self-consistent mean-field-type theories. This is, of course, somewhat unsatisfactory. In other words, neither the size of validity nor the physical relevance of the approximation are not known in the most of self-consistence mean-field-type models. Therefore, we think that a more relevant and more rigorous derivation of a mean-field is still important. That is what we are going to consider in this section. \\

The macroscopic properties of Ising system can be obtained from the following canonical partition function
\begin{equation}
Q=\sum_{\{\sigma\}} e^{- \beta \mathcal{H}\{\sigma\}}, 
\end{equation}
where $ \beta=1/(k_{B}T)  $, $ k_{B}  $ is the Boltzmann's constant and $ T $ is the temperature of the system. For simplicity we display nearest neighbor Ising Hamiltonian for a cubic lattice as 
\begin{eqnarray}
\mathcal{H}\{\sigma_{i,j,k}\}&=-J\sum_{i,j,k}\sigma_{i,j,k}(\sigma_{i+1,j,k}+\sigma_{i,j+1,k}+\sigma_{i,j,k+1}) \nonumber \\
&\qquad  {} {} -h\sum_{i,j,k}\sigma_{i,j,k}, 
\end{eqnarray}
where $ i $, $ j $ and $ k $ show the site of the cubic lattice. The presence of interactions between spins makes the exact treatment of the partition function quite complicated despite the extreme simplicity of the model. It is proven that the exact treatment of this model is very difficult for 2D and 3D systems. Therefore, some approximation is inevitable introduced for the treatment of the system. In most of the previous self-consistent mean-field-type the approximation is made by replacing the central spin with its ensemble average values. As we emphasized in the above, this sort of approximation is not very relevant since it assumes the existence of phase transition from the outset. Therefore, it is instructive to discuss the problem in an alternative way which, though quite crude, does not specifically presume the existance of phase transisiton. Here, we introduce the following approximation for this purpose. We think that it is much more relevant to replace the central spin $ \sigma_{i,j,k} $ with its configurational average instead of replacing it with its ensemble average. The entire discussion of this paper is based on the this configurational consideration. Thus, the configurational average of the central spin can be expressed as
\begin{equation}
\sigma_{i,j,k}=\frac{1}{N}\sum_{i,j,k}\sigma_{i,j,k},
\end{equation} 
where $ N $ is the number of the site of the system. Apparently, the approximation means that the value of the central spin is equal to the sum of the spins in a particular configuration divided by $ N $. At this point we want to mention that the approximation for the central spin is different from Kadanoff's infinite range model \cite{Kadanoff}, where the approximation is introduced to make the coupling between any two spin quite weak. The above consideration for cubic system is readily extended to any hypercubic lattice structures represented by the coordination number $ z $. Thus the Hamiltonian can be expressed as
\begin{equation}
\mathcal{H}\{\sigma_{i,j,k}\}=-J\left( \frac{z}{2}\frac{1}{N}S^{2}\right)-hS, 
\end{equation}
where $ S $ is equal to
\begin{equation}
S=\sum_{i,j,k}\sigma_{i,j,k},
\end{equation}

Before proceeding further, we want to notice that the essence of the mean-field approximation is taken into account within each configuration. But the rest of the calculation over all configuration space is going to be exact. Substituting the above Hamiltonian in the partition function expression leads to 
\begin{equation}
Q=\sum_{\{\sigma\}} e^{\left( \frac{z}{2}\frac{K}{N}S^{2}\right)+HS}, 
\end{equation}
where $ K=J/(k_{B}T)  $ and $ H=h/(k_{B}T)  $. In order to proceed further, the quadratic term in the partition function is expressed with its equivalent form as 
\begin{equation}
e^{\left( \frac{z}{2}\frac{K}{N}\right)S^{2}}=\sqrt{\dfrac{8zK}{\pi N}}\int_{-\infty}^{\infty}e^{-\left(\frac{N}{2zK}\right)t^{2}+St}dt. 
\end{equation}
Using this final Gaussian relation, the partition function turns out to be
\begin{equation}
Q=\sqrt{\dfrac{8zK}{\pi N}}\int_{-\infty}^{\infty}\sum_{\{\sigma\}}e^{-\left(\frac{N}{2zK}\right)t^{2}+S(H+t)}dt.
\end{equation}
The sum over all configuration is easy to compute because of its form as a product of spin terms. Since the sum over $ \sigma $ is 
\begin{equation}
\sum_{\sigma=\pm 1}e^{(H+t)\sigma}=2\cosh(H+t).
\end{equation}
The net result for the partition function is the single integral
\begin{equation}
Q=\sqrt{\dfrac{8zK}{\pi N}}\int_{-\infty}^{\infty}e^{-\left( \frac{N}{2zK}\right)t^{2}} \left[2 \cosh(H+t) \right] ^{N}dt. \label{eq:10}
\end{equation}

The analytical calculation of the integral constitutes thus the essential difficulty in the treatment of our approach. The exact evaluation of this integral, of course, is quite complicated, if not impossible. But it is always possible to develop systematic mathematical approximation for this type calculations. We think that the well-known saddle point method is quite plausible approximation to evaluate the integral. The saddle point approximation procedure in the above integral is mainly based on finding an approximate expression for the $ f(t)=\left[2 \cosh(H+t) \right]^{N} $ function. Let us this function as
\begin{equation}
f(t)= e^{N\ln(2)+N\ln\left[ \cosh(H+t)\right]}.
\end{equation}
Expanding $ \ln\left[ \cosh(H+t)\right]   $ function into Taylor series around $ t_{0}=-H $ produces the relation $ \ln\left[ \cosh(H+t)\right]\simeq \frac{1}{2}(H+t)^{2} $. Thus $ f(t) $ turns out to be
\begin{equation}
f(t)\simeq e^{N\ln(2)+\frac{N}{2}(H+t)^{2}}.
\end{equation} 
Inserting this approximate result into the partition function, it can be written as
\begin{equation}
Q\simeq \sqrt{\dfrac{8zK}{\pi N}}\int_{-\infty}^{\infty}e^{-\left( \frac{N}{2zK}\right)t^{2}+N\ln(2)+\frac{N}{2}(H+t)^{2}}dt.
\end{equation}
The integral can be easily evaluated with Gaussian integral consideration. Then, the result can be expressed as
\begin{equation}
Q\simeq \frac{1}{N}\sqrt{\dfrac{16Kz^{2}}{(1-2z)}} \exp\left[  N\left( \ln(2)+\frac{H^{2}}{2}\right) +\frac{NH^{2}}{ 2\left( \frac{1}{zK}-1\right)}\right].
\end{equation}
The magnetization per spin or the average magnetization $ \langle\sigma\rangle $ is expressed in terms of the partition function as $ \langle\sigma\rangle = \frac{1}{N}\frac{d\ln(Q)}{dH}$. Thus the average magnetization can be easily calculated as
\begin{equation}
\langle\sigma\rangle=H\left[1+\frac{1}{\left( \frac{1}{zK}-1\right)} \right]. 
\end{equation}
To have such a solution where $ \langle\sigma\rangle\neq0 $, it is necessary to consider following heuristic discussion. In the limit $ H $ goes to zero, nonzero values of average magnetization is only possible in the case of $ \frac{1}{zK}-1=0  $. This relation eventually leads to 
\begin{equation}
K_{c}=\frac{1}{z}
\end{equation} 
It is important to notice that this obtained result is equivalent to the well-known result of the Weiss model. The model introduced in this paper, however, yields much valuable physical insight as we have discussed above. \\

We now return to the model introduced this paper. We have seen that it gives qualitatively correct picture of phase transition. However, it cannot produce a relation for the average magnetization for a large range of the critical coupling parameter due to the used saddle point approximation. Therefore, it is inevitable to resort to numerical calculation. Thus, direct evaluation of the derivative $ \frac{1}{N}\frac{d\ln(Q)}{dH} $ in Eq.\eqref{eq:10} easily leads to the following result 
\begin{equation}
\langle\sigma\rangle=\frac{\int_{-\infty}^{\infty}e^{-\left( \frac{N}{2zK}\right)t^{2}}\left[2 \cosh(H+t) \right]^{N}\tanh(H+t)dt}{\int_{-\infty}^{\infty}e^{-\left( \frac{N}{2zK}\right)t^{2}}\left[2 \cosh(H+t) \right]^{N}dt}. \label{eq:17}
\end{equation}

For $ H=0 $, apparently this relation produce $ \langle\sigma\rangle=0 $. One can also find the same result if the invariance of the Ising Hamiltonian is considered. In other words, for $ H=0 $, the contribution of every spin configuration $ \{\sigma\} $  is cancelled by that of $ \{-\sigma\} $. However, in the evaluation of the spontaneous magnetization setting $ H=0 $ is not proper approach. The correct way to calculate the spontaneous magnetization, which is the case of spontaneous symmetry breaking, is to take the thermodynamics limit of Eq.\eqref{eq:17} in the presence of an arbitrarily small $ H $, and then let $ H\rightarrow0 $. Taking into account this discussion, the spontaneous average magnetization can be figured out numerically if $ N $ is large and $ H $ is very small. Of course the true thermodynamics limit is obtained when $ N\rightarrow\infty $ and $ H\rightarrow0 $.  Of course, the values of $ N $ and $ H $ for the numerical calculation are restricted to the available upper limit of computational accuracy. \\
	\begin{figure}[ht]
	\centering
	\includegraphics[width=9cm]{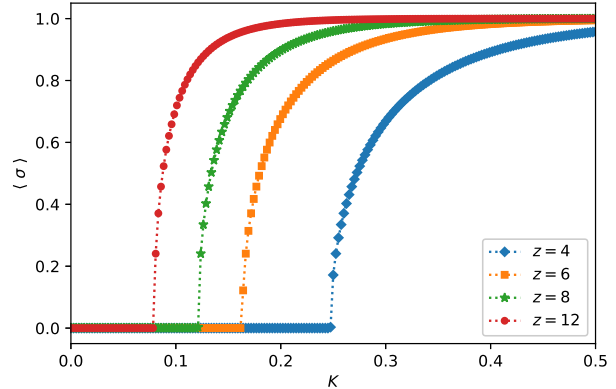}
	\caption{The magnetization per spin $ \langle\sigma\rangle $ with respect to critical coupling strength parameter $ K $ obtained from pure mean-field approach for various hypercubic lattice structures denoted by the coordination number $ z $. The number of spins $ N $ and external magnetic field $ H $ are set to $ 10^{13} $ and $ 10^{-8} $ respectively.}
	\label{fig:1}	
	\end{figure}	

In Fig.\ref{fig:1}, we present the obtained numerical result for various hypercubic lattice structures represented with the coordination number $ z $. In this figure, $ N $ and $ H $ are set to $ 10^{13} $ and $ 10^{-8} $ respectively. The figure indicates that the above obtained analytical result of critical coupling strength parameters are totally in agreement with the numerical results. Therefore it can be claimed that the saddle point approximation is good enough to determine the values of the critical point analytically.\\

As it pointed out earlier, the obtained analytical result of critical point in this section is same as the prediction of the Weiss's mean-field model, namely $ K_{c}=\frac{1}{z} $. This means, the model introduced in this section predicts phase transition for 1D at a finite temperature. Of course, this totally wrong prediction for 1D is the main draw back of the model presented here. In the following section, we consider a more refined method to obtain correct estimation for 1D and also improve the estimation of the critical coupling strengths.   
     
\section{Reduced Transfer Matrix Method}
In this section, we are going to use the reduced transfer matrix method previously introduced \cite{Kayaa,Kayab}. In order to improve the accuracy of estimation of critical coupling strength. In this picture, it is available to handle exactly the interaction between the central spin and one of the its neighbor. In other words, the reduced transfer matrix method retains the interaction along a direction exactly. The interaction between central spin and the rest of the $ \left(\dfrac{z}{2}-1\right)  $ neighboring spins are taken into account with the manner of the above pure mean-field procedure. Thus, the partition function in this scheme can be presented as
\begin{equation}
Q=\sum_{\{\sigma\}}\exp\left[ \sum\sigma_{i,j,k}\sigma_{i+1,j,k}+\left(\frac{z}{2}-1\right)\frac{K}{N}S^{2}+HS\right].     
\end{equation}  

As is in the previous section, we express the quadratic form in the partition function with its equivalent form as 
\begin{equation}
e^{\left(\frac{z}{2}-1\right)\frac{K}{N}S^{2}}=\sqrt{\frac{4(z-2)K}{\pi N}}\int_{-\infty}^{\infty}e^{-\left[ \frac{N}{(z-2)K}\right]t^{2}+St}dt.
\end{equation} 
Using this final relation, the partition function turns out to be
\begin{equation}
Q=A\int_{-\infty}^{\infty}\sum_{\{\sigma\}}e^{K\sum\sigma_{i,j,k}\sigma_{i+1,j,k}-\left[ \frac{N}{(z-2)K}\right]t^{2}+S(H+t)}dt,
\end{equation} 
where  $ A=\left[ \dfrac{4(z-2)K}{\pi N}\right]^{1/2} $. The sum over all configuration is easy to compute because of its similarity to 1D Ising model. Let us write Q in the following form,
\begin{equation}
Q = I \sum_{\{\sigma\}}e^{K\sum  \sigma_{i,j,k}\sigma_{i+1,j,k}+\frac{1}{2}(H+t)\sum ( \sigma_{i,j,k}+\sigma_{i+1,j,k} )}
\end{equation}
where  $ K=J/(k_{B}T)  $ and $ H=h/(k_{B}T) $, and $ I $ is equal to
\begin{equation}
I=\sqrt{\dfrac{4(z-2)K}{\pi N}}\int_{-\infty}^{\infty}e^{-\left[ \frac{N}{(z-2)K}\right]t^{2}}dt.
\end{equation}
Now, the evaluation of the sum over all configurations can be evaluated by the well-known transfer matrix method. Let us a define $ 2\times2 $ matrix $ \textsf{\textbf{P}} $ so that its matrix elements are given by 
\begin{equation}
\langle\sigma_{i,j,k}|\textsf{\textbf{P}}|\sigma_{i+1,j,k}\rangle=e^{K\sigma_{i,j,k}\sigma_{i+1,j,k} +\frac{1}{2}(H+t)(\sigma_{i,j,k}+\sigma_{i+1,j,k})},
\end{equation}
Then $ \textsf{\textbf{P}} $ turns out to be
\begin{equation}
\textsf{\textbf{P}}=\left( {\begin{array}{cc}
	e^{K+(H+t)} & e^{-K} \\
	e^{-K}      & e^{K-(H+t)} \\
	\end{array} } \right).
\end{equation} 
Substituting this matrix into the partition function leads to 
\begin{equation}
Q=I\sum_{\{\sigma\}}\langle\sigma_{1}|\textsf{\textbf{P}}|\sigma_{2}\rangle\langle\sigma_{2}|\textsf{\textbf{P}}|\sigma_{3}\rangle\ldots\langle\sigma_{N}|\textsf{\textbf{P}}|\sigma_{N+1}\rangle .
\end{equation}
If periodic boundry condition $ \sigma_{N+1}=\sigma_{1} $, is assumed, then the partition function turns out to be 
\begin{equation}
Q=I\sum_{\{\sigma\}}\langle\sigma_{1}|\textsf{\textbf{P}}^{N}|\sigma_{1}\rangle=Tr\textsf{\textbf{P}}^{N}=\lambda_{1}^{N}+\lambda_{2}^{N},
\end{equation}
where $ \lambda_{1} $ and $ \lambda_{2} $ are the eigenvalues of the matrix $ \textsf{\textbf{P}} $, which have the following values
\begin{equation}
\lambda_{1,2}=e^{K}\left[\cosh(H+t)\pm\sqrt{\sinh^{2}(H+t)+e^{-4K}} \right]. \label{eq:28}
\end{equation}

Since we are going to work at thermodynamics limits,  $ N\rightarrow\infty $, one can readily conclude that $ \lambda_{1}^{N}\gg\lambda_{2}^{N}  $. Thus, the partition function takes the following form
\begin{equation}
Q=\sqrt{\dfrac{4(z-2)K}{\pi N}}\int_{-\infty}^{\infty}e^{-\left[ \frac{N}{(z-2)K}\right]t^{2}}\lambda_{1}^{N}dt, \label{eq:29}
\end{equation} 
where $ \lambda_{1} $ is as given in Eq.\eqref{eq:28}. The integral can be evaluated approximately by a saddle point calculation as in the previous section. The saddle point approximation procedure in the above integral is mainly based on finding an expression for the $ f(t)=(\cosh(H+t)+[\sinh^{2}(H+t)+e^{-4K}]^{1/2})^{N} $ function. Using the simple fact that $ f(t)=e^{\ln f(t)} $, leads to 
\begin{equation}
f(t)=e^{N\ln\left\lbrace \cosh(H+t)+\left[ \sinh^{2}(H+t)+e^{-4K}\right] ^{1/2}\right\rbrace }.
\end{equation}
Now, expanding the logarithmic function into Taylor series around $ t_{0}=-H $ up to second order term, the result is given by
\begin{equation}
f(t)\simeq e^{N \left[ \ln(1+e^{-2K})+\frac{(1+e^{2K})(H+t)^{2}}{2(1+e^{-2K})}\right] }.
\end{equation}
Substituting this result into the partition function leads to 
\begin{equation}
Q\simeq \sqrt{\dfrac{4(z-2)K}{\pi N}} e^{NK+N\ln(1+e^{-2K})+\frac{NH^{2}}{2(1+e^{-2K})}}\mathcal{G}(t),
\end{equation}
where $ \mathcal{G}(t) $ is as follows
\begin{equation}
\mathcal{G}(t)=\int_{-\infty}^{\infty}e^{-N\left[ \frac{1}{2(z-2)K}-\frac{1}{2e^{-2K}}\right]t^{2}+N\left[ \frac{H}{e^{-2K}}\right]t}dt.
\end{equation}
The expression given above can be calculated by using usual Gaussian integration procedure, then the partition function is expressed as
\begin{equation}
Q\simeq C\sqrt{\dfrac{4(z-2)K}{N^{2}\left(\frac{1}{(z-2)K}-\frac{1}{e^{-2K}} \right) }} \exp\left\lbrace \dfrac{\left( \frac{NH}{^{e-2K}}\right)^{2}}{2N\left[ \frac{1}{(z-2)K}-e^{2K}\right] } \right\rbrace ,
\end{equation}
where, $C= \exp\left[  NK+N\ln(1+e^{-2K})+\frac{NH^{2}}{2(1+e^{-2K})}\right]  $. Now, the magnetization for per spin can be calculated readily from the relation, $ \langle\sigma\rangle = \frac{1}{N}\frac{d\ln(Q)}{dH} $. Evaluation of the derivative leads to 
\begin{equation}
\langle\sigma\rangle \simeq H \left\lbrace  \frac{1}{1+e^{-2K}}+\frac{e^{2K}}{\left[\dfrac{1}{(z-2)K}-e^{2K} \right] } \right\rbrace .
\end{equation}
The critical coupling strength relation corresponding the spontaneous magnetization can be figured out from the above relation as the consideration used in the pure mean-field case in the above section. Thus, the critical coupling strength relation can be obtained as 
\begin{equation}
(z-2)K_{c}e^{2K_{c}}=1.
\end{equation}

Altough the above analytical result obtained with the aid of saddle point approximation is important and valuable, the obtained magnetization relation is not relevant for $ K>K_{c} $. In other words, it has main draw back due to the used saddle point approximation. Therefore, it is necessary to resort to numerical calculations to obtain more proper estimation of the average magnetization quantities. Thus, the direct derivation of the logarithm of the partition function Eq.\eqref{eq:29} with respect to $ H $ leads to   
\begin{equation}
\langle\sigma\rangle=\frac{\int_{-\infty}^{\infty}e^{-\left[ \frac{N}{(z-2)K}\right]t^{2}}\lambda_{1}^{N-1}g(t)dt}{\int_{-\infty}^{\infty}e^{-\left[ \frac{N}{(z-2)K}\right]t^{2}}\lambda_{1}^{N}dt},	
\end{equation}
where, $ \lambda_{1}=e^{K}\left[\cosh(H+t)+\sqrt{\sinh^{2}(H+t)+e^{-4K}} \right] $ and $ g(t)= \sinh(H+t)+\sinh(H+t)\cosh(H+t) \left[\sinh^{2}(H+t)+e^{-4K}\right]^{-1/2} $. \\

\begin{figure}[ht]
	\centering
	\includegraphics[width=9cm]{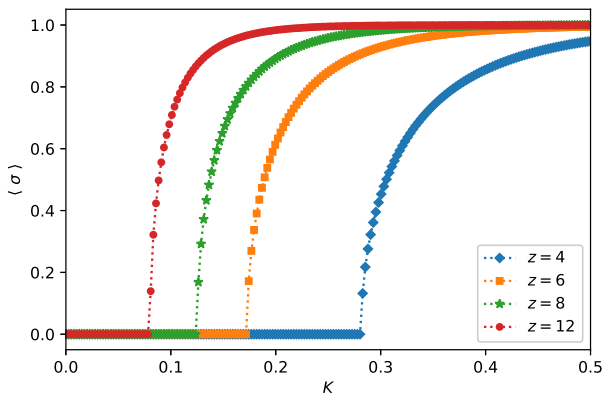}
	\caption{The magnetization per spin $ \langle\sigma\rangle $ with respect to critical coupling strength parameter $ K $ obtained from the reduced transfer matrix method for various hypercubic lattice structures denoted by the coordination number $ z $. The number of spins $ N $ and external magnetic field $ H $ are set to $ 10^{13} $ and $ 10^{-8} $ respectively.}
	\label{fig:2}	
\end{figure}

The relation of average magnetization per spin is solved numerically by setting $ N=10^{13} $ and $ H=10^{-8} $. The datas are plotted in Fig.\ref{fig:2} for various hypercubic lattices. From the figure, it is easily determined that the analytical critical point relation obtained with the aid of saddle point approximation, $ (z-2)K_{c}e^{2K_{c}}=1 $, is totally in agreement with the numerical results. In addition, estimation of the analytical relation for 1D is $ K_{c}\rightarrow\infty $ or $ T_{c}=0 $. Thus, the combination of physical approximation used in this paper and the reduced transfer matrix method produces the exact critical estimation for 1D. In addition, the estimations of critical values for other lattices are better than those of the previous section.

\section{Discussion}
In this paper, we present two different mean-field approaches for Ising model. We mainly focus our attention on physical relevance of the mean-field approximation instead of trying to obtain better estimation of critical points. To do so, we take into account importance of the contribution of each configuration in the evaluation of the average magnetization per spin. Thus, we introduced the approximation used in this paper within each configuration by replacing the central spin with the configurational average. We think that the mean-field approximation used in this paper is physically more relevant and transparent that those used in all the self-consistent mean-field models since they all replace the central spin with the ensemble average of the system. Apparently, the use of ensemble average in the mean-field treatment ignores all the fluctations effects from the outset. In addition, the existence of phase transition is assumed from the outset in any self-consistent mean-field-type models. However, the approximation for central spin used in this paper keeps some of the fluctations implicitly since the evaluation of the partition function is taken all over the possible configurational space. On the other hand, it is often almost trivial to perform the self-consistent mean-field theory calculations. But, analytical calculations of the mean-field theory introduced in this paper is exceedingly more difficult than those of the self-consistent mean-field models. Needless to say that there are many ways to generate mean-field theories. But, a systematic derivation of a mean-field theory is still very important to control the validity of the approximation. Altough the estimated critical values of this paper are less accurate than those of many self-consistent mean-field models, the model of this paper is more systematic and physically more relevant than the many self-consistent mean-field models. Therefore, the relevance of the introduced approximation possible makes the model very useful physical picture for future works. Briefly, the ideas concerned with self-consistent model are interesting and instructive but they do not represent the clearest way of the developing a mean-field model due to their presumption of the existence of phase transition. We have therefore abondened the assumptions of self-consistent mean-field type models in favor of one that emphasizes the essential unity of the treatment and seek to develop physical insight by stressing the configurational content of the approaches.  \\

\end{document}